\documentclass[conference]{IEEEtran}
\IEEEoverridecommandlockouts
\usepackage{cite}
\usepackage{amsmath,amssymb,amsfonts}
\usepackage{algorithmic}
\usepackage{graphicx}
\usepackage{textcomp}
\usepackage{xcolor}
\usepackage{epsfig}
\usepackage{url}

\def\BibTeX{{\rm B\kern-.05em{\sc i\kern-.025em b}\kern-.08em
    T\kern-.1667em\lower.7ex\hbox{E}\kern-.125emX}}

\begin{document}

\title{Performance of an Astrophysical Radiation Hydrodynamics Code
under Scalable Vector Extension Optimization\\
}

\author{\IEEEauthorblockN{Dennis C.\ Smolarski}
\IEEEauthorblockA{\textit{Department of Mathematics \& Computer Science} \\
\textit{Santa Clara University}\\
Santa Clara, USA\\
dsmolarski@scu.edu}
\and
\IEEEauthorblockN{F.\ Douglas Swesty, Alan C. Calder}
\IEEEauthorblockA{\textit{Department of Physics and Astronomy} \\
\textit{Stony Brook University}\\
Stony Brook, USA \\
\{douglas.swesty, alan.calder\}@stonybrook.edu}
}

\maketitle

\begin{abstract}
We present results of a performance study of an astrophysical radiation hydrodynamics code, V2D,  on the Arm-based A64FX processor developed by Fujitsu. The code solves sparse linear systems, a task for which the A64FX architecture should be well suited.  We performed the performance analysis study on Ookami, an Apollo 80 platform utilizing the A64FX processor.   We explored several compilers and performance analysis packages and found the code did not perform as expected under scalable vector extension optimization, suggesting that
a ``deeper dive" into analyzing the code is worthwhile.  However, a simple driver program that exercised basic sparse linear algebra routines used by V2D did show significant speedup with the use of the scalable vector extension optimization. We present the initial results from the study which used V2D on a relatively simple test problem that emphasized the repeated solution of sparse linear systems.
\end{abstract}


\section{Introduction}

\subsection{Astrophysical Radiation Transport}
Many astrophysical simulations require modeling the flow of some form of radiation, either photons or neutrinos, through gas.  Modeling the flow of radiation requires the parallel, implicit solution of a large, sparse, set of coupled linear equations that describe the time evolution of the radiation energy density across a spectrum of energies.   In this paper we investigate the computational performance of one 
such simulation code, V2D~\cite{SwestyMyra2005}, which models
the flow of radiation in the multigroup flux-limited diffusion approximation. V2D uses a parallelized Krylov subspace algorithm to solve the linear system of equations arising from the finite-difference discretization of the underlying equations that describe the diffusive evolution of the radiation energy density.

\subsection{The Ookami Platform}
Ookami~\cite{ookamiurl} is a test bed supported by the United States
National Science Foundation (NSF).  The platform is an HPE Apollo 80 with 
174 A64FX Fujitsu compute nodes, each with 32GB high-bandwidth memory 
and a 512GB SSD. 
The A64FX processors consist of four core memory groups each with 12 cores,
64KB L1 cache, and 8MB L2 cache shared between the cores and runs at 1.8
GHz. The processor uses the Armv8.2--A Scalable Vector Extension (SVE)
SIMD instruction set with 512 bit vector implementation. This allows for
vector lengths anywhere from 128--2048 bits and enables vector length
agnostic (VLA) programming. Ookami has an
Infiniband HDR100 fat tree interconnect with 200 gigabit switches, and
a high-performance Lustre file system provides about 800 TB storage.

Ookami was the first open machine outside of Japan featuring
the A64FX processor, and the aim of the Ookami project is 
to provide researchers with
access to this state-of-the-art scientific computing technology 
in order to explore it and demonstrate 
its potential. 
The code, V2D, is one of the application codes
being ported and tuned for the machine. More details on the system 
and the project can be found in \cite{pearc_experiences_2021,ookamied2021}.
Ookami is presently administered by Stony 
Brook University and the University at Buffalo, but in October 
2022 Ookami will be an XSEDE level 2 service provider
\cite{ookamixsede}.

\subsection{The V2D Code}
The V2D code uses finite-difference algorithms to solve the
equations of Eulerian hydrodynamics and multi-species 
flux-limited diffusive radiation transport in two spatial dimensions.    It was designed primarily for the purpose of simulating 
core collapse supernovae but has a wider applicability to many radiation-hydrodynamic problems.   Details related to the underlying numerical methods can be found in \cite{SwestyMyra2005}.  Written in Fortran-95, V2D employs MPI for domain-decomposed parallelism and HDF5 for parallel input and output.  The linear system solver uses a restructured version of the BiCGSTAB \cite{vdv92} 
algorithm, which gangs inner products to reduce the number of parallel global reduction operations required per iteration of the BiCGSTAB solver.   Preconditioning of the linear system is accomplished using a sparse approximate inverse preconditioner \cite{SSS2004}.   Because of its prohibitive size,  the sparse linear system matrix is never stored and the Krylov subspace methods are implemented in matrix-free form by application of a finite-difference operator to
column vectors that are stored as Fortran arrays defined with the same spatial shape as the 2D grid.   This strategy also avoids the costly packing/unpacking of data into some form of sparse matrix storage each time a linear system must be solved.
The V2D code has been generically written to allow various coordinate systems and the $x1$ and $x2$ spatial directions are always considered to
be orthogonal.  The problem is domain decomposed using a Cartesian 2-D spatial tile decomposition that is controlled by adjustable run-time parameters {\tt NPRX1} and {\tt NPRX2} that control the tiles in the $x1$ and $x2$ directions, respectively.
Thus the process topology may be varied to better apportion the load among processors. Finally, we
note that the calculations tested here were performed in double precision arithmetic.

\section{The SVE Optimization Study}

\subsection{The V2D Radiation Test Problem}
The test diffusive radiation transport problem that we consider in this paper involves the diffusion of a 2-D Gaussian pulse of radiation\cite{SwestyMyra2005} and does not involve hydrodynamic evolution.    This particular test problem was chosen for this study because the principal computational effort is expended in the solution of a large, sparse, memory-bandwidth-limited linear system that describes the time evolution of the radiation distribution.   Solving this system should be a task for which the Ookami architecture is ideally suited.    The linear system, as described below, consists of $x1 \times x2 \times 2$ coupled linear equations, where the spatial dimensions are $x1 = 200$ and $x2= 100$ zones respectively, and the number of radiation species is 2.   We wish to emphasize that this is a small test problem that we have chosen for a study of performance under SVE optimization. In the work in this paper we make only limited use of the parallel capability of V2D when we vary the process topology to adjust the problem size on each processor.

\begin{figure}[h] \centerline{ \epsfxsize=3.60in
\epsffile{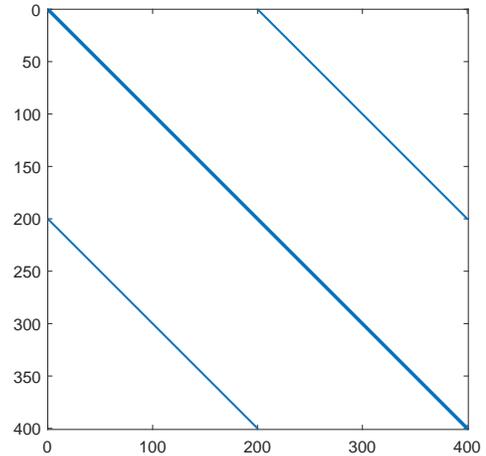}} \caption{The sparsity pattern of matrices
used in the V2D code.} \label{fig1} \end{figure}

While the aforementioned linear system is sparse, it has a regular structure. If the matrix corresponding to this system were actually stored, with a dictionary ordering it would form a banded matrix with five bands.   A portion of what the sparsity pattern for this matrix would look like is depicted in Fig. \ref{fig1}.
The figure only depicts the upper left 400 $\times$ 400 block of the complete 40,000 $\times$ 40,000 matrix. On either side of the diagonal are two adjacent diagonals with two outlying diagonals spaced farther from the diagonal.  The $x1$ parameter indicates the distance of the two outlying diagonals from the center diagonal.      

The Krylov subspace algorithms employed by V2D require only matrix-vector multiplication operations (hereafter Matvec), and not the matrix itself.  
Because the matrix would arise from a second-order spatial finite-difference scheme for the diffusion operator, we instead evaluate the Matvec operations by applying the diffusion operator in finite-difference form to a vector to form a new resultant vector.   For each element of the vector there are relatively few floating-point operations needed to evaluate the Matvec operation and thus the Matvec operations are memory bandwidth limited.

The Krylov subspace algorithm we employ for the linear system solution is the BiCGSTAB algorithm\cite{vdv92}.   BiCGSTAB is an extension of the Conjugate Gradient (CG) method (which is designed for a symmetric linear system of the form $Ax=b$, where $A$ would represent the system matrix) to those cases  where the system matrix $A$ is non-symmetric.   The time evolution of the 
radiation energy density in this test problem involves many repeated solutions of linear systems, one for each timestep, thus causing the computational cost of the simulation to be dominated by the sparse matrix operations employed by the BiCGSTAB algorithm.

\subsection{Compilers and SVE Optimization}
Our first task was to identify the prerequisites of V2D, e.g.,
MPI~\cite{MPI:Forum,Dongarra1995AnIT} and HDF5~\cite{hdf5-website},
and determine how those are accessed and configured on Ookami.
We then tested combinations of compilers and MPI implementations.  As a sample
of results, those presented below indicate test runs using the GNU (ver. 11.1.0), Fujitsu (ver. 4.5),
and Cray (ver 21.03) compilers.  Some compilers allowed the use of either MVAPICH
~\cite{mvapich}
or OpenMPI~\cite{openmpi} and others only provided one option for MPI.

\subsection{Performance Monitoring}

We compared the CPU times of simulations compiled with different compilers, both with and without SVE optimization. 
Because the {\it craypat} performance tool will only work with
the Cray compiler, the
Linux {\tt perf stat} command of the Linux kernel performance
monitor, with the {\tt -e duration\_time} and 
{\tt -e cpu-cycles} flags, was used to measure the time of the simulations for the different compilers. This command measures the entire CPU time of the process.
Each configuration (of the total number of processors used and the process topology in the $x$1- and $x$2- directions, determining how the linear system was partitioned) was run several times to confirm the timing results.

In addition, for the fastest versions of the executable, produced by the Cray compiler, the TAU software 
(developed at the University of Oregon \cite{Tau})
was also used to verify the times recorded from the runs. TAU's ParaProf visualization tool also enabled us to see which routines contributed most to the total time without the need to add additional routine calls in the software.

\subsection{Results for SVE Optimization of V2D}

Table \ref{table1}
summarizes the results of using three different
compilers on Ookami to produce a V2D executable that solves a linear system consisting of $x1 \times x2
\times 2$ equations, where $x1 = 200$ and $x2= 100$.  
The test problem time--evolves the radiation 
energy density for 100 time steps.   Each time step requires the solution of three unique $x_1\times x_2 \times 2$ linear systems via the BiCGSTAB algorithm.   Thus this test problem establishes performance results for the solution of 300 linear systems.
As noted above, the times were clocked using the Linux {\tt perf stat} 
performance analysis command,
which was used to measure the duration of the execution, 
giving values in seconds.

All the compilers available on Ookami are able to make use of the SVE
capabilities of the architecture.  In the trials listed in Table \ref{table1}, the
SVE and optimization features were used, but we also turned off the SVE and other optimization features
on some of the Cray trials for comparison purposes.

In Table \ref{table1} the values in the $N_p$ column refer to the total number of processors used for a run, 
with the values in the ``Direction" {\tt NX1} column indicating the number of domain decomposition tiles
in the $x$1 direction and similarly for the values in the {\tt NX2} column. 
Thus the product of the two values equals the total number of processors requested.  The column labeled Cray (opt) indicates results obtained with an executable
compiled both with both {\tt -O3} optimization and SVE optimization enabled.  The last column, labeled ``Cray (no-opt)", used an executable compiled with the Cray compiler without either the {\tt -O3 } optimization or SVE optimization enabled.

\begin{table}
\begin{center}
\caption{Times by compiler\label{table1}}
\begin{tabular}{|c|r|r|r|r|r|r|}
\hline
\textbf{$N_p$}&\multicolumn{2}{|c|}{\textbf{Direction}} &
\multicolumn{4}{|c|}{\textbf{Times by Compiler (seconds)}} \\ \hline
      &  \multicolumn{1}{|c|}{\tt NX1} &  \multicolumn{1}{|c|}{\tt NX2} &  \multicolumn{1}{|c|}{GNU} &  \multicolumn{1}{|c|}{Fujitsu} & \multicolumn{1}{|c|}{Cray} & \multicolumn{1}{|c|}{Cray} \\ 
    & & &  &  & \multicolumn{1}{|c|}{(opt)} & \multicolumn{1}{|c|}{(no-opt)}\\ \hline \hline
	1	&	1	&	1	&	363.91  & 252.31 & 181.26 &262.57\\ \hline
	10	&	10	&	1	&	  43.85 & 31.76 & 24.20 & 32.35\\ \hline
	20	&	20	&	1	&	  26.80 & 19.79 & 16.78 & 20.66\\ \hline
	20	&	10	&	2	&	  25.74 & 19.66 & 15.73& 19.93\\ \hline
	20	&	5	&	4	&	  25.42 & 18.85 & 15.39& 19.79\\ \hline
	25	&	25	&	1	&	  24.62 & 17.24 & 15.65&\\ \hline
	40	&	40	&	1	&	  25.30 & 13.97 & 19.12&\\ \hline
	40	&	20	&	2	&	  22.88 & 12.96 & 17.37&\\ \hline
	40	&	10	&	4	&	  21.91 & 13.04 & 17.16&\\ \hline
	50	&	50	&	1	&	  30.10 & 13.05 & 25.56&\\ \hline
	50	&	25	&	2	&	  29.26 & 12.09 & 24.07&\\ \hline
	50	&	10	&	5	&	  27.55 & 11.40 & 23.51& \\ \hline
\end{tabular}

\end{center}
\end{table}

\subsection{Timing analysis}

The timing information in Table \ref{table1}
reveals that the Cray and Fujitsu compilers produced substantially faster code than the GNU compiler.   The Cray compiler also produced slightly faster code than the Fujitsu compiler when using 25 or fewer processors.

The various trials using the Cray compiler with and without SVE optimization reveal a reduction in CPU time with
the use of SVE optimization.  Using PAPI analysis calls, we noted speedup using SVE optimization in the routines that applied the preconditioner to the system
matrix, as well as when using the dot product, vector addition, and
combined vector scaling/addition routines.

When using a single processor, the majority of time was spent in the
matrix-vector multiplications, approximately 141 seconds out of 181,
with preconditioning taking about 14 additional seconds. We also make use of Arm's MAP performance analysis tool, which indicated that the three calls to the BiCGSTAB routine each took approximately 31-33\% of the total time using a single processor.

When using 20 processors, in a $5\times 4$ configuration, approximately
7.5 seconds out of 15 were spent in the matrix-vector multiplications
at maximum per processor, with preconditioning taking about 0.8 seconds
at maximum. As to be expected with multiple processors, a significant
amount of time was taken by MPI calls.

Increasing the number of processors will decrease the amount of time
needed for the matrix-vector products, but there is a point, varying with each compiler, at which the increased time needed for inter-processor communication via MPI calls does not
lead to further overall reduction in time.  The variations exhibited in
Table \ref{table1} between different compilers can be attributed to the
machine architecture used for vector multiplications as well as the
different ways the compilers optimize the code.

These results were qualitatively as expected---optimization with SVE produced
a speedup. But the magnitudes of the speedups observed with the Cray
compiler were smaller than what we expected for memory-bound 
linear system operations.  We attributed this to the overall complexity of the multi-physics V2D code. V2D was designed with abstracted operators for linear algebra, but calls to these operators are
interspersed with calls to other physics routines and a simulation exercises more than just the operators. Further analysis is certainly needed.

\subsection{Timing V2D matrix operations using PAPI} 

Because the use of SVE optimization did not produce the speedup in V2D that we had hoped for, we therefore wrote a simple single-processor driver program that exercised the actual V2D routines that are utilized in the BiCGSTAB solver without the added complications of the other V2D code. 
With this driver program we then examined the CPU time utilized for repeated computation of the
linear algebra routines, both with and without the SVE optimization enabled.
We used a linear system with 1000 equations and repeated
operations 100,000 times.  The CPU times spent in the linear-algebra routines were obtained both from checking
the hardware clock and by using PAPI software timers, but the differences between the two
were insignificant.

As Table \ref{table2} indicates, when these V2D sparse linear-algebra routines were exercised using the driver program, a significant speedup was realized with SVE optimization.
Overall, the use of SVE (with the Cray compiler) reduced the time spent in these routines to
approximately between 16\% and 31\% of the time needed without SVE.
(The times are given in seconds in the following table.)  MATVEC = Matrix
Vector Product; DPROD = Dot Product; DAXPY = computation of $a\cdot x
+ y$; DSCAL = computation of $c - d\cdot y$; DDAXPY = computation of
$a\cdot x + b\cdot y + z $.

\begin{table}
\caption{Linear algebra routines times \label{table2}}
\begin{center}
    \begin{tabular}{|c|c|c|c|}\hline
\multicolumn{4}{|c|}{\textbf{PAPI times (seconds)}}\\ \hline
    \textbf{Routine} & \textbf{No-SVE} & \textbf{SVE} & \textbf{SVE/No-SVE} \\ \hline \hline
    MATVEC & 599 & 96 & 0.16 \\
    DPROD & 132 & 24.3 & 0.18 \\
    DAXPY & 206 & 53.8 & 0.26 \\
    DSCAL & 153 & 47.7 & 0.31 \\
    DDAXPY & 296 & 65 & 0.22 \\ \hline 
    \end{tabular}
\end{center}
\end{table}

\section{Conclusions and Future Work}

The principal conclusion, which in hindsight we realize is fairly obvious, is that a complex multi-physics code, even though it is dominated by memory bandwidth-limited sparse linear algebra computations, will not necessarily demonstrate the speedup expected with the use of SVE optimization.   However,  
testing just the memory bandwidth-limited sparse linear 
algebra routines did reveal that they were able to undergo significant speedup with SVE optimization.  
Because V2D relies on these same routines for its implementation of the BiCGSTAB algorithm, we need to do further work to understand what is limiting the overall V2D code performance.   Future work will entail
more detailed instrumentation of the code to delineate the origins of the performance bottleneck.

As also noted, the times recorded when using the Fujitsu compiler with more than 25 nodes were less than times recorded with the Cray compiler. Further investigation with a larger problem and more nodes comparing the Fujitsu and Cray compilers is warranted, along with the use of other compilers, such as Clang~\cite{clang}.
\newpage

%

\section*{Acknowledgment}

The authors would like to thank Stony Brook Research Computing and 
Cyberinfrastructure, and the Institute for Advanced Computational Science 
at Stony Brook University for 
access to the innovative high-performance Ookami 
computing system, which was made possible by a \$5M National Science Foundation 
grant (\#1927880). This research was supported in 
part by the US Department of Energy (DOE) under grant DE-FG02-87ER40317.
The authors gratefully acknowledge the Ookami support team and the other
users who cheerfully answered questions and offered assistance during
this study. Without their invaluable help this work would have been
extremely difficult, if not impossible. 


\bibliographystyle{IEEEtran}


\end{document}